\documentclass[10pt]{article}
\usepackage{fancyhdr}
\usepackage{extramarks}
\usepackage{amsmath}
\usepackage{amsthm}
\usepackage{amsfonts}
\usepackage{siunitx}
\usepackage{tikz}
\usepackage[plain]{algorithm}
\usepackage{algpseudocode}
\usepackage{multirow}
\usepackage{booktabs}
\usepackage{graphicx}
\usepackage{subfigure}
\usepackage[colorlinks,linkcolor=black,anchorcolor=black,citecolor=black,urlcolor=blue]{hyperref}
\usepackage{amsmath,bm}
\usepackage{booktabs}
\usepackage{mathtools}
\usepackage{amssymb}
\usepackage{caption}
\usepackage{capt-of}
\usepackage{mciteplus}
\usepackage{cite}
\usepackage{mathrsfs}
\usepackage[title,titletoc,toc]{appendix}
\usepackage{xr}
\usepackage{parskip}
\usepackage{soul}
\usepackage{textcomp}
\usepackage[colaction]{multicol}
\usepackage[switch]{lineno}
\usepackage{lipsum}
\usepackage{etoolbox}
\usepackage{longtable}
\usepackage{array}
\usepackage{tablefootnote}
\usepackage{multirow}
\newcolumntype{C}[1]{>{\centering\arraybackslash}p{#1}}
\usetikzlibrary{automata,positioning}
\usetikzlibrary{automata,positioning}

\usepackage{ulem}
\usepackage{xcolor,tikz}

\begin{document}

\title{Mathematical artificial intelligence  design of  mutation-proof COVID-19 monoclonal antibodies}

\author{Jiahui Chen$^1$ 
 and Guo-Wei Wei$^{1,2,3}$\footnote{
Corresponding author.		E-mail: weig@msu.edu} \\
$^1$ Department of Mathematics, \\
Michigan State University, MI 48824, USA.\\
$^2$ Department of Electrical and Computer Engineering,\\
Michigan State University, MI 48824, USA. \\
$^3$ Department of Biochemistry and Molecular Biology,\\
Michigan State University, MI 48824, USA. \\
}
\date{\today} 

\maketitle


\begin{abstract}
     Emerging severe acute respiratory syndrome coronavirus 2 (SARS-CoV-2) variants have compromised existing vaccines and posed a grand challenge to coronavirus disease 2019 (COVID-19) prevention, control, and global economic recovery. For COVID-19 patients, one of the most effective COVID-19 medications is monoclonal antibody (mAb) therapies. The United States  Food and Drug Administration (U.S. FDA) has given the emergency use authorization (EUA) to a few mAbs, including those from Regeneron, Eli Elly, etc. However, they are also undermined by SARS-CoV-2 mutations. It is imperative to develop effective mutation-proof mAbs for treating COVID-19 patients infected by all emerging variants and/or the original SARS-CoV-2. We carry out a deep mutational scanning to present the blueprint of such mAbs using algebraic topology and artificial intelligence (AI). To reduce the risk of clinical trial-related failure, we select five mAbs either with FDA EUA or in clinical trials as our starting point. We demonstrate that topological AI-designed mAbs are effective to variants of concerns and variants of interest designated by the World Health Organization (WHO), as well as the original SARS-CoV-2. Our topological AI methodologies have been validated by tens of thousands of deep mutational data and their predictions have been confirmed by results from tens of experimental laboratories and population-level statistics of genome isolates from hundreds of thousands of patients.
\end{abstract}


\section{Introduction}

In combating  the coronavirus disease 2019 (COVID-19) pandemic, there has been exigency to develop effective antiviral treatments i.e., vaccines, antiviral drugs, and antibody therapies. The developments of these treatments are some of the most paramount scientific accomplishments in  the battle against COVID-19. However, emerging severe acute respiratory syndrome coronavirus 2 (SARS-CoV-2) variants, particularly variants of concern (VOCs),  impact transmission, virulence, and immunity and pose a threat to existing vaccines and antibody drugs.

SARS-CoV-2 is an enveloped, unsegmented positive-sense single-strand ribonucleic acid (RNA) virus, which enters cells depending on the binding of its spike (S) protein receptor-binding domain (RBD) to host angiotensin-converting enzyme 2 (ACE2) receptor\cite{hoffmann2020sars}. The binding free energy (BFE) between the S protein and ACE2, according to epidemiological and biochemical analysis, is proportional to the infectivity of  SARS-CoV-2   in the host cells \cite{qu2005identification,wang2021analysis}. In July 2020, it was shown that  driven by natural selection \cite{chen2020mutations}, mutations strengthen RBD-ACE2 binding  and thus make the virus more  infectious.  The high-frequency  RBD mutations were shown to be undoubtedly governed by  natural selection \cite{chen2020mutations,wang2021vaccine}. Additionally, natural selection also creates new SARS-CoV-2 variants easily escaping antibodies induced  by either infection or vaccination \cite{wang2021mechanisms}. By comparing to the first SARS-CoV-2 strain deposited to GenBank (Access number: NC 045512.2), the mutation-induced BFE changes ($\Delta\Delta G$) of the binding of S protein and ACE2 provide a way to measure the infectivity changes of a SARS-CoV-2 variant. Positive BFE changes induced by mutations of RBD binding to ACE2 reveal that mutations potentially improve the binding, while negative BFE changes indicate mutations weaken the transmissibility and infectivity. Thus, the impact of SARS-CoV-2 RBD variants on infectivity can be evaluated according to their BFE changes \cite{li2005bats,chen2021prediction,chen2020mutations,chen2021revealing}.

Currently, except for antiviral drugs which are proved more efficacious than placebo such as Pfizer's Paxlovid (nirmatrelvir),
COVID-19 vaccines are considered as the game-changer and SARS-CoV-2 monoclonal antibody (mAb) therapies are shown to reduce the risk of disease progression. Both approaches rely on antibodies in different mechanisms. Specifically, vaccines are designed to stimulate an effective host immune response triggering the host adaptive immune system to produce  antibodies against future infection \cite{amanat2020sars}, while antibody therapies are obtained from patients convalescing from COVID-19 or other diseases, which block viral entry by binding to the viral S protein. Various vaccines, including two mRNA vaccines designed by Pfizer-BioNTech and Moderna, have been granted authorization for emergency use as well as antibody therapies (such as casirivimab\cite{hansen2020studies}, imdevimab\cite{hansen2020studies}, bamlanivimab\cite{jones2020ly}, etesevimab\cite{shi2020human}, regdanvimab\cite{kim2021therapeutic},  et al.) in many countries. However, RBD mutations simultaneously  strengthen SARS-CoV-2 infectious \cite{chen2020mutations}, escape existing vaccines \cite{wang2021mechanisms}, and attenuate antibodies \cite{wang2021antibody}.

Genetic mutations of SARS-CoV-2 provide a mechanism for viruses to adapt to and evade host immune responses, COVID-19 vaccines, and antibody therapies. Although SARS-CoV-2 has a higher fidelity and a slower evolutionary rate than other RNA viruses \cite{day2020evolutionary}, over 5,000 unique mutations were found on SARS-CoV-2 S protein \cite{wang2021vaccine,chen2021prediction}. This situation awakes the question of the impacts of existing mutations on vaccines and antibodies. According to the WHO tracking SARS-CoV-2 variants \cite{trackingWHOvariants}, variants are characterized as Variants of Interest (VOIs) and Variants of Concern (VOCs) and prioritized for global monitoring and research. Other variants of local interest/concern are designated by national authorities. There are more than ten designated VOCs, including Alpha (B.1.1.7), Beta (B.1.351), Gamma (P.1),  Delta (B.1.617.2),  etc. It is interesting to note that RBD residues 452 and 501 were predicted to  ``have high changes to mutate into significantly more infectious COVID-19 strains''  in early 2020 \cite{chen2020mutations}. As predicted,  variants Alpha, Beta, Gamma, Delta, Kappa, Theta, Lambda, Mu, etc. all have at least one of these two mutations. 

Evidence shows VOCs have high transmissibility and dominate the spreading of SARS-CoV-2 on multiple countries \cite{wang2021antibody,davies2021estimated,planas2021reduced,planas2021reduced} (see Fig.~\ref{fig_combine_all}a). Studies show VOCs are resistant to antibody neutralization. For example, Alpha and Beta variants are reported as antibody resistance  to neutralization by some anti-N-terminal domain (NTD) and anti-BRD mAbs,   including casirivimab and bamlanivimab for Beta variants\cite{wang2021antibody}. Gamma variant is also shown refractory to neutralization by some mAbs, including emergency use authorization (EUA) antibody therapies casirivimab, imdevimab, and etesevimab \cite{wang2021increased,jangra2021sars}, and similar results are for Delta variant as well \cite{planas2021reduced}.  Additionally, according to WHO \cite{trackingWHOvariants},  VOIs including Eta, Iota, Kappa, and Lambda variants have genetic changes impacting virus characteristics of transmissibility, disease severity, immune escape, and diagnostic escape and lead to significant community transmission. VOIs may share some mutations with VOCs on RBD. Thus, single mutation experiments on L452R, S477N, and E484K, can be used to analyze their effects on antibody neutralization  \cite{liu2021identification,jangra2021sars,deng2021transmission}. For example, the mutation L452R on the S protein RBD increases 20\% of the transmissibility of SARS-CoV-2\cite{deng2021transmission}, and has mild negative impacts on the neutralization by EUA antibody therapies according to Food and Drug Administration (FDA)  \cite{eua1,eua2}.

Experimental studies of mutational impacts on the existing antibodies and antibody drugs are time-consuming and are limited to a small fraction of known viral mutations. It is difficult to accurately determine whether a mutation will evade a vaccine in general populations of various races, genders, ages, and existing health conditions. Based on the molecular mechanism of host cells infected by SARS-CoV-2 virus and immune system responses, quantitative assessment of mutational  impacts on SARS-CoV-2 infectivity and  antibody drugs can be achieved by computing BFE changes following mutations of the RBD-ACE2 complex and  RBD-antibody complexes. In our earlier work, we applied a topology-based deep learning model to predict the binding free energy (BFE) changes of the RBD-ACE2 complex and 106 RBD-antibody complexes induced by RBD mutations\cite{chen2020mutations,chen2021prediction,wang2021vaccine,chen2021revealing}. These  predictions were validated by experimental results \cite{chan2020engineering,starr2020deep,linsky2020novo,starr2021prospective,greaney2021complete}. For example, our predictions of mutation-induced BFE changes on CTC-445.2 binding to RBD were shown to be highly correlated with the experimental data \cite{linsky2020novo,chen2021prediction}. In recent work, we validated our predictions of BFE changes on the RBD-ACE2 complex with deep mutational scanning data, achieving the Pearson correlation of 70\% \cite{linsky2020novo,chen2021prediction}. Moreover, in a comparison with experimental data, the predicted BFE changes have an 80\% correlation  with the escape fraction \cite{chen2021prediction}.  A high prediction accuracy with experimental data was found in predicting emerging variant impacts on clinical trial antibodies \cite{chen2021prediction}.

The objective of this work is to introduce a mathematical  artificial intelligence (AI)-based  computational strategy for the rational design of mutation-proof mAbs. As examples, we consider high-frequency RBD mutations on 5 mAb therapies, namely casirivimab, imdevimab, etesevimab, bamlanivimab, and regdanvimab. Among them,  casirivimab and imdevimab are authorized  for the treatment of COVID-19 by the U.S. Food and Drug Administration (FDA).   Etesevimab and bamlanivimab are also obtained  FDA's emergency use authorization (EUA).  Regdanvimab is issued advice on use for treating COVID-19 by European Medicines Agency (EMA). We use our intensively-validated algebraic topology-based deep learning model to estimate the mutation-induced BFE changes of antibody-RBD complexes. This study also offers an important strategy for the design of mutation-proof mAbs for other viruses. 

\section{Results}

\subsection{AI-based deep mutational screening of five RBD-binding antibodies}
\begin{table}[ht!]
	\centering
	\caption{The statistics of BFE changes  ($\Delta\Delta G$) induced by AI-based deep mutations on antibody variable domains. The number of AI-based deep mutations on each chain is denoted as ``Total No''. Three categories of the numbers of topological AI-based mutations are given to BFE changes greater than 0 kcal/mol, 0.5 kcal/mol, and 1 kcal/mol, respectively.}
	\label{tab_all}
	\begin{tabular}{|cc|c|cc|cc|cc|}
		\hline
		\multirow{2}{*}{Antibody}&\multirow{2}{*}{Chain} &\multirow{2}{*}{Total No} & \multicolumn{2}{c|}{$\Delta\Delta G>0$ kcal/mol} & \multicolumn{2}{c|}{$\Delta\Delta G>0.5$ kcal/mol} & \multicolumn{2}{c|}{$\Delta\Delta G>1$ kcal/mol}\\
		& &  & No & Ratio (\%) & No & Ratio (\%) & No & Ratio (\%) \\\hline
		\multirow{2}{*}{REGN10933}&Heavy& 2223 & 742 & 33.38 &46 & 2.07 & 19 & 0.85\\
		&Light& 1995 & 858 & 43.01 & 11 & 0.55 & 1 & 0.05\\\hline
		\multirow{2}{*}{REGN10987}&Heavy& 2223 & 675 & 30.36 & 24 & 1.08 & 11 & 0.49\\
		&Light& 1995 & 734 & 36.79 & 7 & 0.35 & 1 & 0.05 \\\hline
		\multirow{2}{*}{LY-CoV016}&Heavy& 2242 & 220 & 9.81 & 8 & 0.36 & 2 & 0.09 \\
		&Light& 2090 & 168 & 8.04 & 2 & 0.10 & 1 & 0.05 \\\hline
		\multirow{2}{*}{LY-CoV555}&Heavy& 2337 & 480 & 20.54 & 35 & 1.50 & 5 & 0.21 \\
		&Light& 2014 & 518 & 25.72 & 11 & 0.55 & 3 & 0.15 \\\hline
		\multirow{2}{*}{CT-P59}&Heavy& 2394 & 514 & 21.47 & 18 & 0.75 & 8 & 0.33 \\
		&Light& 2090 & 542 & 25.93 & 9 & 0.43 & 0 & 0.00 \\\hline
		\multicolumn{2}{|c|}{Average} & 2160 & 545 & 25.51 & 17 & 0.77 &5 & 0.23 \\\hline
	\end{tabular}
\end{table}
We first carry out  a topological AI-based deep mutational scanning on the antibody variable domains that bind to the RBD for five mAbs. These mutations are conducted systematically such that each residue in each mAb's light and heavy chains is mutated to all 19 other possible amino acids. Then, the BFE change for the antibody-RBD complex induced by each mutation is computed by the topological AI model. Most mutations on the antibody variable domain tend to have negative BFE changes or mild positive BFE changes (see supplementary information), indicating that mAbs have been optimized for their RBD binding.  Table~\ref{tab_all} shows the statistical results for five mAbs involving about 21,600 AI-based deep mutations on antibody variable domains. An average of 25.51\% mutations cause the strengthening of antibody-RBD binding (or having positive BFE changes). In fact, only  0.77\% and 0.23\% mutations have BFE changes greater than 0.5 kcal/mol and 1 kcal/mol, respectively. The dramatic decrease in the number of  mutations having BFE changes greater than 0.5 kcal/mol indicates that these antibodies have a small number of residue sites for improving the mAb neutralization effect against SARS-CoV-2. Among the five antibodies, LY-CoV016 has the least number of antibody mutations for strengthening its binding with RBD, while REGN10933 has a relatively large number of residues that can be improved.
The heap map of complete virtual mutational scans on the antibody variable domains is provided in the Appendix. 

In Figure~\ref{fig_combine_all}c, the residues with at least one mutation having BFE changes greater than 1 kcal/mol are presented according to Table~\ref{tab_all}. For REGN10933, two residues A75 and T102 on the heavy chain have four mutations (A75Y/W/F/M) and seven mutations (T102D/E/Q/W/I/L/V) with BFE changes greater than 1 kcal/mol. 
For the heavy chain of REGN10987, A33 has eight candidates (A33K/D/E/Q/T/I/L/M) for strengthening the binding of REGN10987 and RBD. For the rest of selected residues, none of them have more than three effective mutants. These small numbers of candidates also indicate that these antibody therapies were optimized. However, their optimizations were respect  to the original SARS-CoV-2 virus and these mAbs are prone to emerging RBD mutations.

\subsection{AI-based rational design of mutation-proof antibodies}

SARS-CoV-2 variants have been evolving to increase their capability to evade vaccine and antibody protections \cite{wang2021mechanisms}. With the threat of emerging SARS-CoV-2 variants, it is important to design mutation-proof antibody therapies. 
Our essential idea is to systematically mutate each residue of an antibody into 19 possible other amino acids to search for mutation-proof new designs of antibodies.  
Variants Alpha (B.1.1.7), Beta (B.1.351), Gamma (P.1), Delta (B.1.617.2), Lambda (C.37), Epsilon (B.1.427), and Kappa (B.1.427) encode spike proteins with mutations K417N/T, L452R/Q, T478K, E484K/Q, F490S, and N501Y in the spike protein RBD that provide a degree of resistance to neutralization by our previous modeling prediction \cite{chen2021revealing} and experimental analysis \cite{tada2021neutralization,wu2019large,grubaugh2020making,garcia2021multiple,wibmer2021sars,wu2021serum,tada2021sars} (see Fig. \ref{fig_combine_all}b). In addition to WHO designated variants, the 10 most observed RBD mutations in terms of their frequencies are more infectious and increase the virus transmissibility \cite{chen2021revealing}, which include seven mutations appearing in the WHO designated variants plus S477N, N439K, and S494P. Mutation S477N, N439K, and S494K rank 5th, 7th, and 9th in terms of frequencies. Mutations L452Q and E484Q of Lambda and Kappa variants, respectively, where E484Q ranks 11th, are not in the top ten observed RBD mutations \cite{wang2021vaccine}. Thus, we focus on these twelve mutations for the antibody redesigning and provide the 100 most observed RBD mutation results in the Appendix.

\subsubsection{REGN10933 and REGN10987}

\begin{figure}[htb]
	\setlength{\unitlength}{1cm}
	\begin{center}
	\includegraphics[width=6.3in]{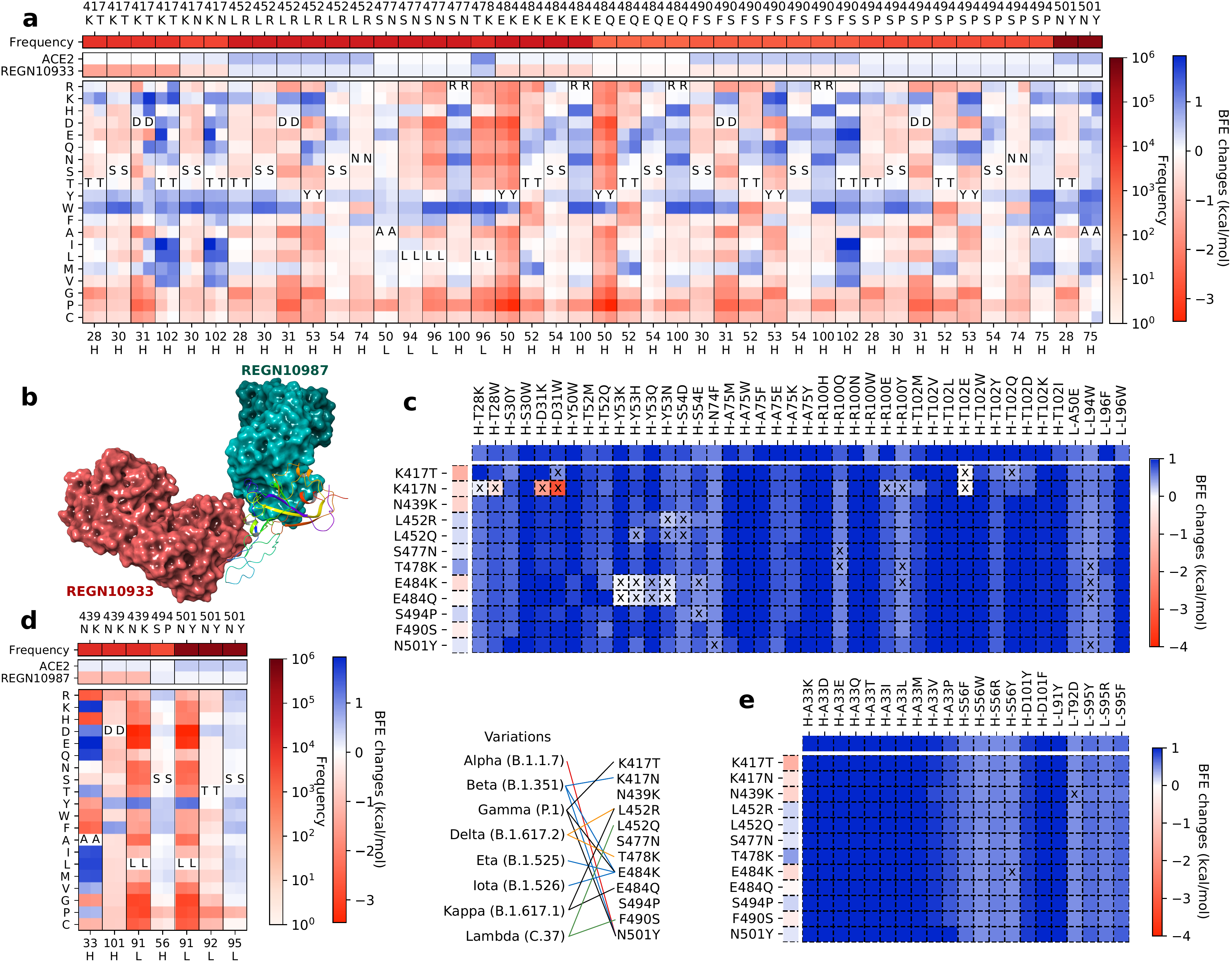}
	\caption{The deep mutational analysis on antibodies REGN10933 and REGN10987. {\bf a} The mutational scanning on antibody REGN10933 binding to S protein RBD and mutated  RBD. In the bottom column labels, H indicates the heavy chain of REGN10933 and L indicates the light chain. {BFE change range is from -3.46kcal/mol to 1.94kcal/mol.} {\bf b} 3D structure of the complex (PDB: 6XDG)\cite{hansen2020studies}. {\bf c} Illustration of BFE changes of effective antibody mutations. The first column indicates the BFE changes induced by RBD mutations of the binding between RBD and antibody. The first row indicates the BFE changes induced by antibody mutations of the binding between RBD and antibody. {\bf d} The mutation scanning on antibody REGN10987 binding to S protein RBD and mutated RBD. {\bf e} Illustration of BFE changes of effective antibody mutations.}
	\label{fig_RN33}
	\end{center}
\end{figure}
As shown  in Figures~\ref{fig_RN33}a and~\ref{fig_RN33}d, the analysis of antibodies REGN10933 and REGN10987 are given for the deep mutational scanning on antibody variable domains that bind to the original S protein RBD and mutated RBD of variants.
The mutations on antibodies are considered if the distances between C$\alpha$s of antibody residues and RBD residues are less than 15 {\AA} and selected when antibody mutations have positive BFE changes greater than 0.5 kcal/mol both for binding to the original RBD and the RBD of variants. 
Figure~\ref{fig_RN33}a shows ten mutations on the S protein RBD with effective mutations on antibody REGN10933. For unselected RBD mutations N439K and L452Q, the deep mutational scanning on antibody variable domains within 15 {\AA} to the RBD shows no BFE changes greater than 0.5 kcal/mol. The first row of Figure~\ref{fig_RN33}a gives the frequency information of each RBD mutation. The following two rows give the BFE changes following the RBD mutations of the binding between S protein RBD and ACE2 or between the RBD and antibodies, where RBD mutations are more favorable of the binding to ACE2 than to antibody REGN10933. The rest rows demonstrate the deep mutational scanning on REGN10933 binding to S protein RBD on odd columns and RBD mutations on even columns. For a pair of the RBD and its mutation, there are multiple residues on antibody REGN10933 having mutations that increase the binding affinity for both. Notice that not all mutations on residues have positive BFE changes, and in total, there are 42 candidates on antibody REGN10933.

Once the 42 candidates of antibody REGN10933 are selected, their BFE changes of the binding to the RBD with the 12 mutations induced by antibody mutations are displayed in Figure~\ref{fig_RN33}c. The cross marks indicate that the BFE changes are less than 0.5 kcal/mol. Meanwhile, the first column of the heatmap gives the BFE changes of the RBD binding to antibody induced by RBD mutations, and the first row of the heatmap gives the BFE changes of the binding complex induced by antibody mutations. Here, 17 of 42 mutations on REGN10933 have one or more BFE changes less than 0.5 kcal/mol. Especially, D31W on the heavy chain of REGN10933 causes a negative BFE change of -3.16 kcal/mol. Y53N on the heavy chain and L94W on light chain induces four BFE changes less than 0.5 kcal/mol. For antibody REGN10933, mutations H-S30Y/W, H-Y50W, H-T52Q, H-A75M/W/F/E/K/Y, H-R100H/N/W, H-T102M/V/L/W/Y/D/K/I, L-A50E, and L-L96F/W can be the effective candidates for improving the neutralization of antibody REGN10933 against the S protein RBD and its variants.

\begin{figure}[htb]
	\setlength{\unitlength}{1cm}
	\begin{center}
		\includegraphics[width=5.8in]{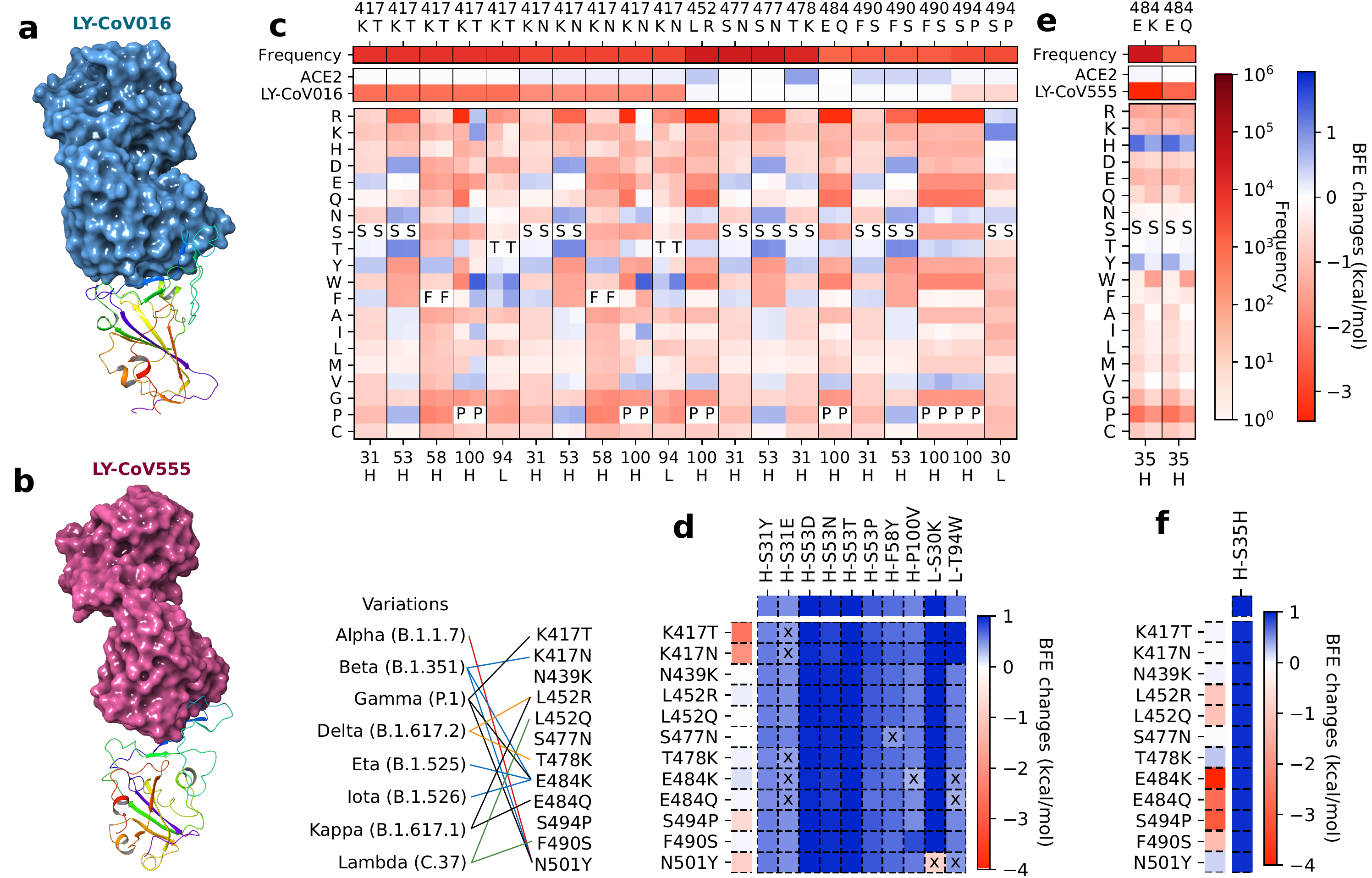}
		\caption{The deep mutational analysis on antibodies LY-CoV016 and LY-CoV555. {\bf a} 3D structure of LY-CoV016 and RBD (PDB: 7C01)\cite{shi2020human}. {\bf b} 3D structure of LY-CoV555 and RBD (PDB: 7KMG)\cite{jones2020ly}. {\bf c} The deep mutational scanning on antibody LY-CoV016 binding to S protein RBD and mutated  RBD. In the bottom column labels, H indicates the heavy chain of LY-CoV016 and L indicates the light chain. {BFE change range is from -3.46kcal/mol to 1.94kcal/mol.} {\bf d} Illustration of BFE changes of effective antibody mutations. The first column indicates the BFE changes induced by RBD mutations of the binding between RBD and antibody. The first row indicates the BFE changes induced by antibody mutations of the binding between RBD and antibody. {\bf e} The deep mutational scanning on antibody LY-CoV555 binding to S protein RBD and mutated RBD. {\bf f} Illustration of BFE changes of effective antibody LY-CoV55 mutations.}
		\label{fig_LY_CoV}
	\end{center}
\end{figure}

For antibody REGN10987, there are 21 candidates on the variable domain and 10 candidates are on the heavy chain A33 in Figure ~\ref{fig_RN33}e. Considering REGN10987 not directly connecting S protein RBD on the receptor binding motif of ACE2, three RBD mutations are selected with distances less than 15 {\AA} to REGN10987 (see Fig.~\ref{fig_RN33}b) and having BFE changes greater than 0.5 kcal/mol for both cases (see Fig.~\ref{fig_RN33}d). Except for the antibody mutations with BFE changes less than 0.5 kcal/mol, there are 19 mutations H-A33K/D/E/Q/T/I/L/M/V/P, H-S56F/W/R/Y, H-D101Y/F, L-L91Y, L-T92D, and L-S95Y/R/F that improve the neutralizing efficacy against SARS-CoV-2 and variants.

\subsubsection{LY-CoV016 and LY-CoV555}

With a similar analysis on antibodies LY-CoV016 (see Fig.~\ref{fig_LY_CoV}a) and LY-CoV555 (see Fig.~\ref{fig_LY_CoV}b), we collected 10 candidates and one candidate for LY-CoV016 and LY-CoV555, respectively, in Figure~\ref{fig_LY_CoV}. In Figure~\ref{fig_LY_CoV}c, seven RBD mutations are evaluated on 6 antibody residues H-S31, H-S53, H-F58, H-P100, L-S30, and L-T94. Interestingly, half of the residues are serine, which has a small polar uncharged side chain. From the second and third column of Figure~\ref{fig_LY_CoV}c, it is noticed that RBD mutations are more favorable to the RBD binding to ACE2 than to LY-CoV016. Eliminating the mutations with BFE changes less than 0.5 kcal/mol, there are five effective mutations H-S31Y and H-S53D/N/T/P for LY-CoV016 for improving its competitiveness (see Fig.~\ref{fig_LY_CoV}d). As for LY-CoV555, only one candidate H-S35H is selected as shown in Figures~\ref{fig_LY_CoV}e and f. 

\subsubsection{CT-P59}

\begin{figure}[htb]
\setlength{\unitlength}{1cm}
\begin{center}
	\includegraphics[width=6.5in]{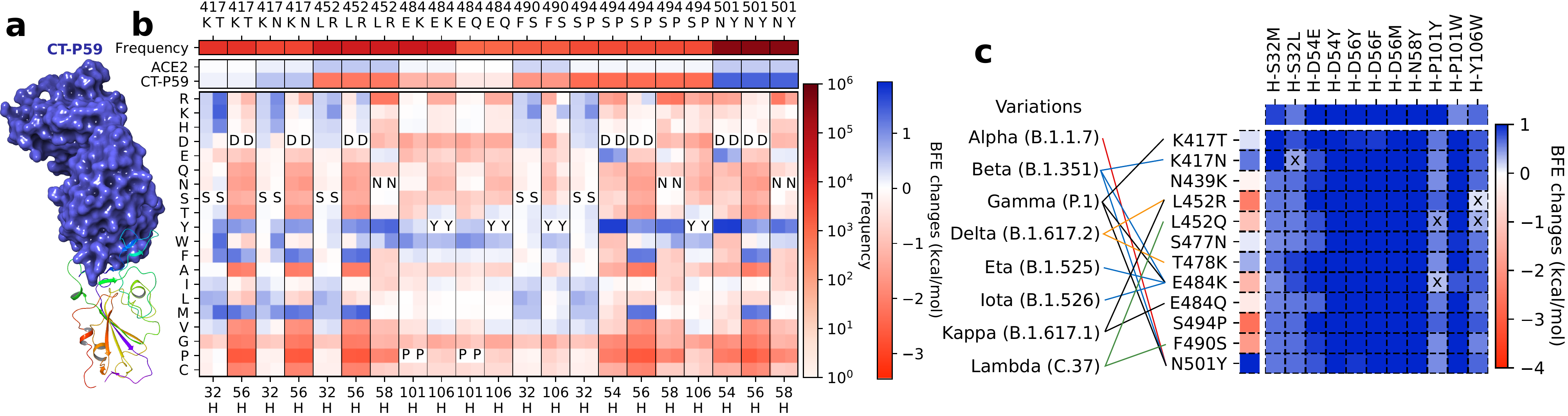}
	\caption{The deep mutational analysis on antibody CT-P59. {\bf a} 3D structure of CT-P59 and RBD (PDB: 7CM4)\cite{kim2021therapeutic}. {\bf b} The mutational scanning on antibody CT-P59 binding to S protein RBD and mutated  RBD. In the bottom column labels, H indicates the heavy chain of CT-P59. {BFE change range is from -3.46kcal/mol to 1.94kcal/mol.} {\bf c} Illustration of BFE changes of effective antibody mutations. The first column indicates the BFE changes induced by RBD mutations of the binding between RBD and antibody. The first row indicates the BFE changes induced by antibody mutations of the binding between RBD and antibody.}
	\label{fig_CT-P59}
\end{center}
\end{figure}

Finally, we analyze antibody CT-P59 with 11 effective mutations on 6 residues on the heavy chain, which are H-S32M/L, H-D54E/Y, H-D56Y/F/M, H-N58Y, H-P101Y/W, and H-Y106W (see Fig.~\ref{fig_CT-P59}). There are seven RBD mutations K417T/N, L452R, E484K/Q, F490S, S494P, and N501Y. For example, mutation L452R, a mutation of the Delta variant, is favorable to the neutralization of binding to ACE2, but disrupting the neutralization of the binding to CT-P59. Six candidates H-S32M/L, H-N56Y/F/M, and H-N58Y on CT-P59 can counteract the disrupting effect by mutation L452R. In addition, the RBD mutation T478K, another mutation of the Delta variant, has a mild positive BFE change of its binding to CT-P59 and positive BFE changes of its binding to CT-P59 with mutations as shown in Figure~\ref{fig_CT-P59}c. Although the Delta variant reduces susceptibility against CT-P59, the selected candidates H-S32M/L, H-D54E/Y, H-D56Y/F/M, H-N58Y, and H-P101W can improve the neutralizing efficacy of CT-P59 against the Delta variant.
Notably, our early predictions of BFE changes induced by mutations L452R and T478K binding to CT-P59\cite{chen2021revealing} are proved by later experimental results on the neutralization of CT-P59 binding to the single mutants of the RBD \cite{lee2021therapeutic}. Overall, mutations H-S32M/L, H-D54E/Y, H-D56Y/F/M, H-N58Y, and H-P101W on CT-P59 can improve the neutralization ability of CT-P59 binding to the S protein RBD. 

\section{Discussion}
\begin{figure}[htb]
	\setlength{\unitlength}{1cm}
	\begin{center}
		\includegraphics[width=6.5in]{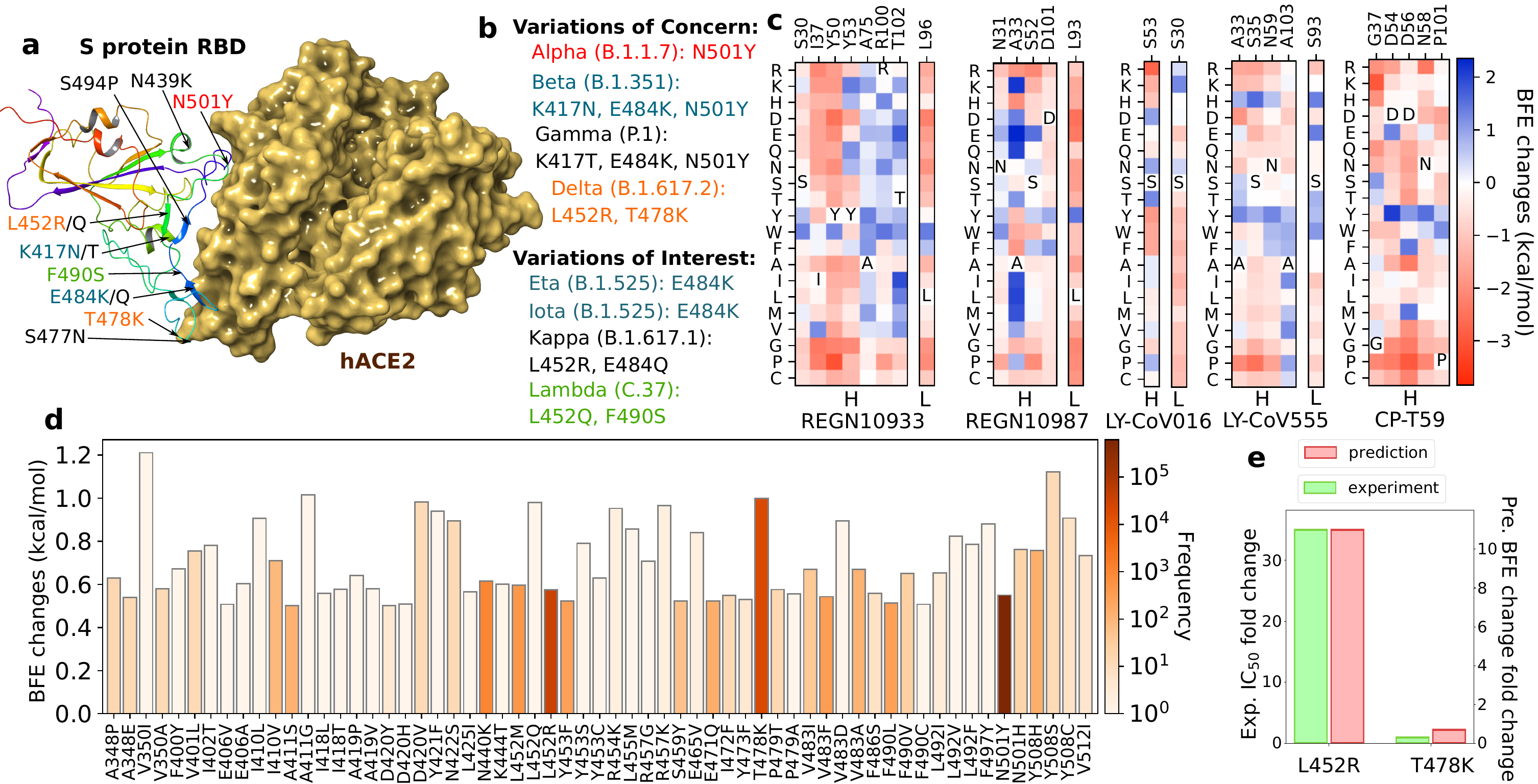}
		\caption{{\bf a} 3D structure of host ACE2 (ACE2) and RBD (PDB: 6M0J)\cite{lan2020structure}. {\bf b} The lists of variants of concerns (VOCs) and variants of interest (VOIs). {\bf c} Residues on antibodies selected with at least one mutation having BFE changes greater than 1 kcal/mol. {BFE change range is from -3.84kcal/mol to 2.35kcal/mol.} {\bf d} Illustration of SARS-CoV-2 mutation-induced BFE changes for the complexes of S protein and ACE2. Here, mutations with BFE changes greater than 0.5 kcal/mol are presented. {\bf e} Comparison of experimental fold change and predicted fold change in IC$_{50}$ of CT-P59 and S protein RBD complex induced by mutations L452R and T478K.}
		\label{fig_combine_all}
	\end{center}
\end{figure}
Emerging variants have dominated the spreading of  SARS-CoV-2 worldwide   and have been shown to reduce the neutralization efficacy of antibodies and degrade the protection of SARS-CoV-2 vaccine and antibody treatments. Especially, more attention should be paid to RBD mutations as the S protein RBD and ACE2 binding is the key for SARS-CoV-2 virus host cell entry. The SARS-CoV-2 RBD mutations can strengthen the RBD-ACE2 binding to make the virus more infectious and meanwhile, weaken the RBD-antibody binding to breakthrough vaccines and mAbs.  Consequently, the efficacy of vaccines and antibody therapies are compromised and viral transmissibility is enhanced. Twelve mutations on RBD are observed from variants of concern (VOCs) and variants of interest (VOIs) (see Fig. \ref{fig_combine_all}). Interestingly, according to our prediction on BFE changes induced by RBD mutations for RBD binding to human ACE2, all these VOCs have at least one mutation on RBD which has the BFE change greater than 0.5 kcal/mol. For instance, Alpha, Beta, and Gamma variants have the mutation N501Y with a BFE change 0.55 kcal/mol, and the Delta variant has mutations L452R and T478K with BFE changes 0.58 kcal/mol and 1.00 kcal/mol, respectively. The BFE strengthening   mutations on the Delta variant RBD enhance the infectivity of the Delta variant, creating a dominant strain. For VOIs, none of them have a mutation on RBD with BFE changes greater than 0.5 kcal/mol. The highest BFE change is induced by L452Q from the Lambda variant that is 0.44 kcal/mol. Thereafter, the hypothesis is that emerging SARS-CoV-2 variants have at least one mutation with BFE changes greater than 0.5 kcal/mol. Based on our previous findings in \cite{chen2020mutations}, 606 out of 1149 RBD mutations that we predicted as ``most likely'' mutations have been observed, while the rest mutations 1912 ``likely'' and 625 ``unlikely'' mutations are rarely found on the S protein RBD. In Figure \ref{fig_combine_all}, we list 61 most likely mutations on RBD whose BFE changes are greater than 0.5 kcal/mol, 38 of 61 mutations have been observed and 17 mutations V350I, I410L, A411G, D420V, Y421F, N422S, L452Q, R454K, L455M, R457K, E465V, T478K, V483D, L492V, F497Y, Y508S/C have BFE changes from 0.82 kcal/mol to 1.21 kcal/mol,  which could be effective mutations for VOCs. Note that a high BFE change of the binding between S protein RBD and ACE2 indicates the strengthening of SARS-CoV-2 infectivity. Potentially, this mutation could be a vaccine escape mutation if it weakens the binding of RBD to antibodies.

\section{Validation}

The validation of our topological AI model predictions for mutation-induced BFE changes  has been demonstrated by comparison with experimental data in recent publications \cite{chen2021prediction,chen2021revealing}. Firstly, we showed high correlations of experimental deep mutation enrichment data and predictions for SARS-CoV-2 S protein RBD and CTC-445.2 complex\cite{chen2021prediction} and SARS-CoV-2 RBD and ACE2 complex\cite{chen2021revealing}. In the comparison with experimental data on clinical trial antibody therapies for high-frequency mutations, our predictions achieve a Pearson correlation of 0.80 \cite{chen2021revealing}. Considering the BFE changes induced by mutations on the RBD of the ACE2-RBD complex, predictions on mutations L452R and N501Y have a highly similar trend with experimental data \cite{chen2021revealing}. Meanwhile, as we presented early  \cite{wang2021vaccine}, high-frequency mutations are associated with positive BFE changes. Moreover, for multi-mutation tests, our BFE change predictions have the same pattern with experimental data of the impact of SARS-CoV-2 variants on major antibody therapeutic candidates \cite{chen2021revealing}.

Recent studies on the potency of CT-P59 in vitro and in vivo against Delta variants \cite{lee2021therapeutic} show that the neutralization of CT-P59 is reduced by effects of L452R (13.22 ng/mL) and is retained against T478K (0.213 ng/mL). In our predictions\cite{chen2021revealing}, L452R induces a negative BFE change (-2.39 kcal/mol) and T478K induces a positive BFE change (0.36 kcal/mol). In Figure~\ref{fig_combine_all}e, the fold changes are presented for experimental and prediction values. 

Further validation on the Alpha variant RBD mutation was discussed elsewhere \cite{wang2022emerging}. Our predictions of Omicron BA.1 and BA.2 infectivity, vaccine breakthrough, and antibody resistance, which were made when there were no experimental results available, were later nearly perfectly confirmed by experimental data \cite{chen2022omicron, chen2022omicron2}.

\section{Methods}

The development of our deep learning model for BFE change predictions on protein-protein interactions for SARS-CoV-2 problems can be summarized in four steps.  First, preparing genome sequence data from the GISAID database \cite{shu2017gisaid} (\url{https://www.gisaid.org/}). By taking the first complete SARS-CoV-2 genome from the GenBank (NC\_045512.2) as the referencing \cite{wu2020new}, a set of single nucleotide polymorphism (SNP) profiles is generated, i.e., residues 329 to 530 on the S protein RBD have 606 non-degenerate mutations are found. Then, 100 most observed mutations have been collected with frequency more than 40 times. Next, collecting SARS-CoV-2 data and related data is the key step, which makes the model reliable and accurate. Massive data of BFE changes of SARS-CoV-2 are rarely reported, while the enrichment ratio data via high-throughput deep mutations are relatively easy to obtain. With the fundamental dataset of BFE changes upon mutations the SKEMPI 2.0 dataset \cite{jankauskaite2019skempi}, deep mutational enrichment ratio data is added as another database for our machine learning training \cite{chen2021revealing}. After the database preparation, the third step is the feature generations of protein-protein interaction complexes. We implemented the element-specific algebraic topological analysis on point cloud samples consisting of complex atoms \cite{wang2020topology, chen2021revealing,cang2018representability}. This topological approach is based on persistent homology \cite{zomorodian2005computing, edelsbrunner2008persistent}, a powerful method for protein structure representation \cite{cang2018representability,xia2014persistent} and drug discovery \cite{nguyen2020review}.  Additionally, biophysics and biochemistry features such as surface areas, partial charges, Coulomb interactions, et al., are combined with topological features\cite{chen2021prediction}. Lastly, deep neural networks are constructed for the BFE change prediction of protein-protein interactions involving mutations \cite{chen2021revealing}.

In the third step, obtaining the mutant protein structure requires using Scap utility from Jackal software package\cite{xiang2001extending}, which replaces the side chain of the mutation site with min option being set to 4 with additional conformers obtained by perturbing conformers in a rotamer library. The mutant protein structures of RBD variants are constructed by Scap and, then, are used as primal structures for the calculation of antibody mutation impact on RBD variants. 

The detailed descriptions of datasets and machine learning model are given in literature\cite{chen2020mutations,wang2020mutations,chen2021revealing} and are available at \href{https://github.com/WeilabMSU/TopNetmAb}{TopNetmAb}. In addition,
the SARS-CoV-2 single nucleotide polymorphism data in the world is available at \href{https://users.math.msu.edu/users/weig/SARS-CoV-2_Mutation_Tracker.html}{Mutation Tracker}. 
The analysis of RBD mutations is available at \href{https://weilab.math.msu.edu/MutationAnalyzer/}{Mutation Analyzer}.

\section{Conclusion}

Driven by natural selection \cite{chen2020mutations}, severe acute respiratory syndrome coronavirus 2 (SARS-CoV-2) has been evolving towards increasingly more infectious, vaccine escape, and antibody resistance \cite{wang2021mechanisms}. Interestingly, this evolution can be achieved through mutations at the viral spike protein receptor-binding domain (RBD), which binds to the human angiotensin-converting enzyme 2 (ACE2) to facilitate the viral cell entry. Meanwhile, the RBD is also a target of most monoclonal antibodies (mAbs) for direct neutralization of the virus. As a result, natural selection-driven virus evolution gives rise to variants of concerns (VOCs), such as Variants Alpha, Beta, Gamma, Delta, etc. VOCs fuel the waves of widespread infections, evade vaccines, and attenuate the efficacy of existing mAbs. This work provides a mathematical  artificial intelligence (AI)-based strategy to design mutation-proof antibodies  

Our mathematical  AI model utilizes persistent homology and deep learning and was trained with tens of thousands of experimental data, including SARS-CoV-2 related deep mutational data. We carry out an AI-based deep mutational screen of five existing mAbs, including those approved by the U.S. Food and Drug Administration (FDA) for emergency use authorization (EUA). Our deep mutational screen indicates that most mAbs have been optimized against the original SARS-CoV-2 but are prone to the RBD mutations. By considering high-frequency RBD mutations, including those from VOCs, we systematically mutate each residue of the five selected mAbs to 19 possible variants to search for potentially mutation-proof new mAbs. Our study offers many alternative designs of mutation-proof mAbs.

\section*{Acknowledgment}
This work was supported in part by NIH grant  GM126189, NSF grants DMS-2052983,  DMS-1761320, and IIS-1900473,  NASA grant 80NSSC21M0023,  Michigan Economic Development Corporation, MSU Foundation,  Bristol-Myers Squibb 65109, and Pfizer.
The authors thank The IBM TJ Watson Research Center, The COVID-19 High Performance Computing Consortium, NVIDIA,  and MSU HPCC for computational assistance.
The authors thank Drs. Kaifu Gao and Changchuan Yin, and Ms. Rui Wang for useful discussion.

\section*{Appendix}
\begin{figure}[h!]
	\setlength{\unitlength}{1cm}
	\begin{center}
		\includegraphics[width=6.5in]{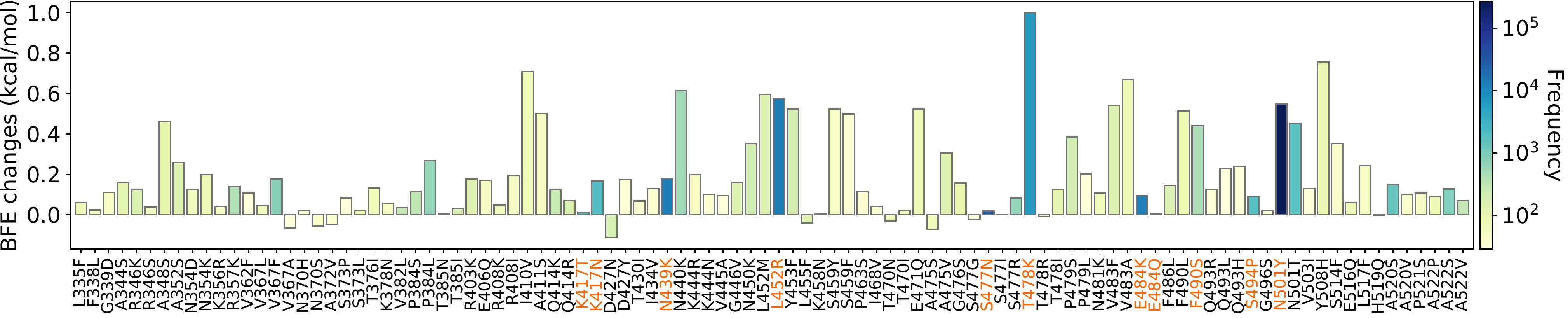}
		\caption{BEF changes of the 100 most observed RBD mutations with their frequency. The variants' mutations are colored in orange. }
		\label{fig_top100}
	\end{center}
\end{figure}
The 100 most observed RBD mutations are collected with their BFE changes and frequency correspondingly. In Figure~\ref{fig_top100}, variants' mutations are colored in orange. Mutations L452R, T478K, and N501Y are predicted with high BFE changes.

\begin{figure}[h!]
	\setlength{\unitlength}{1cm}
	\begin{center}
		\includegraphics[width=6.5in]{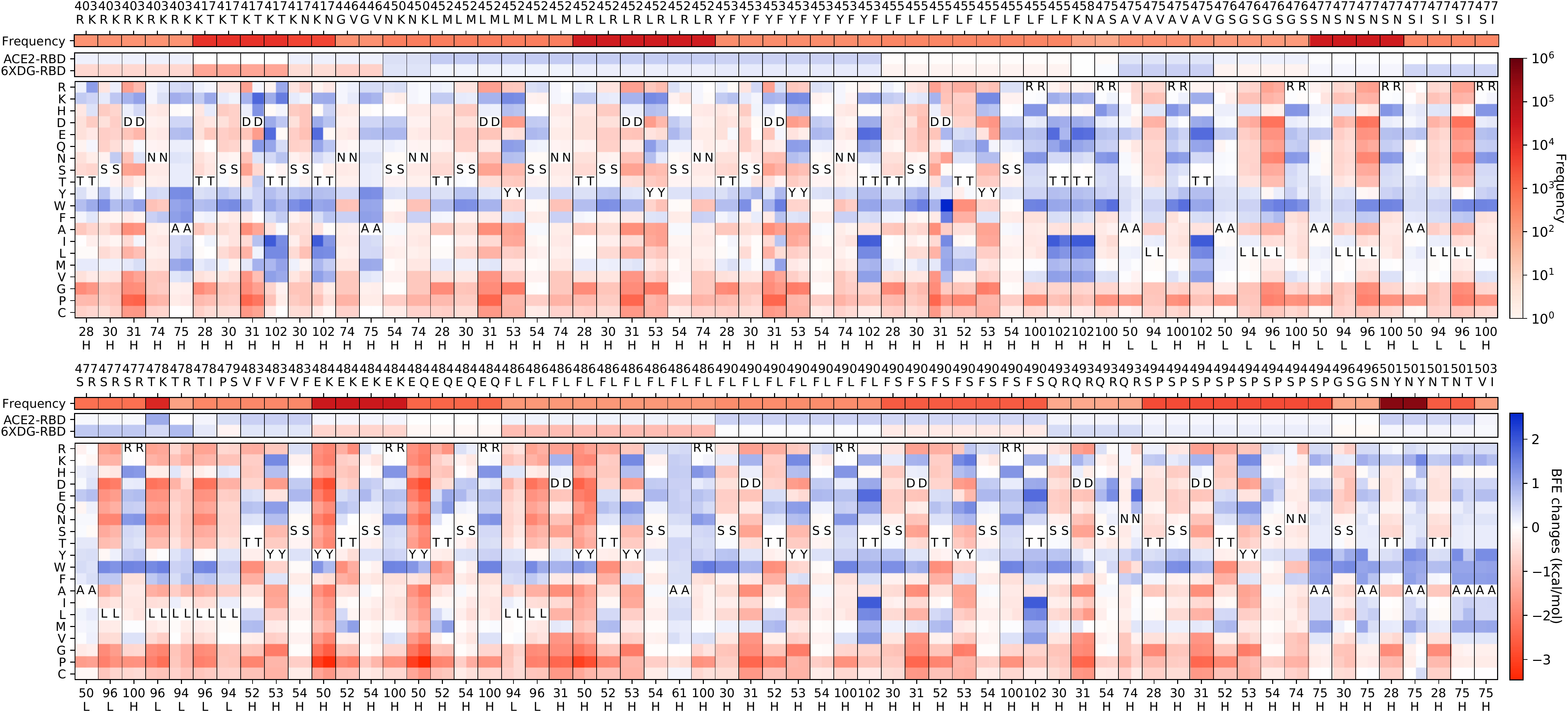}
		\caption{The mutational scanning on antibody REGN10933 binding to S protein RBD and mutated  RBD. In the bottom column labels, H indicates the heavy chain of REGN10933. L indicates the light chain of RENG10933. {BFE change range is from -3.28kcal/mol to 2.41kcal/mol.}}
		\label{fig_RN33_all}
	\end{center}
\end{figure}
\begin{figure}[h!]
\setlength{\unitlength}{1cm}
\begin{center}
	\includegraphics[width=6.5in]{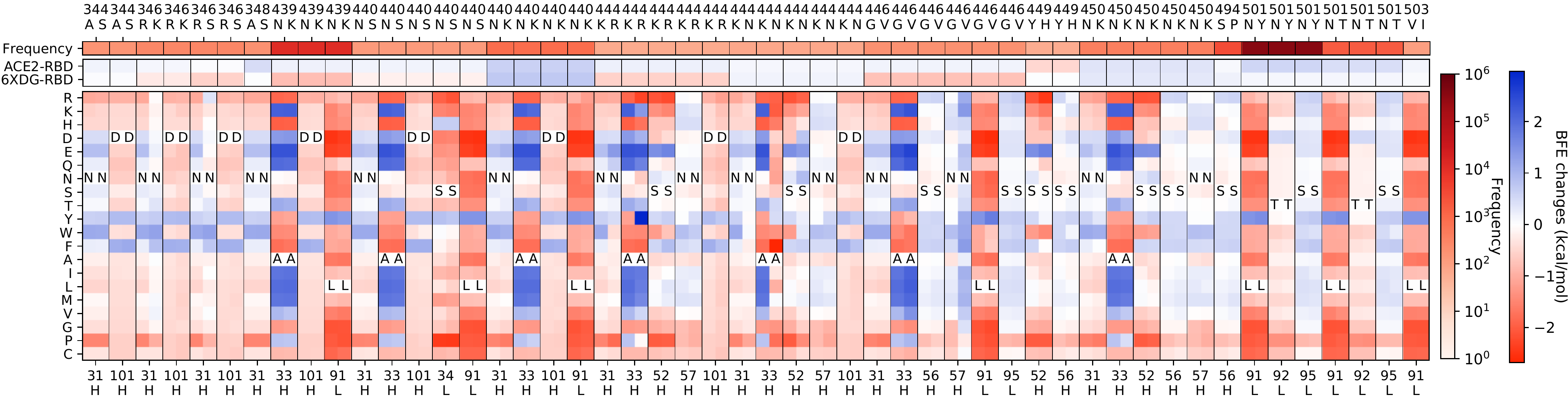}
	\caption{The mutational scanning on antibody REGN10987 binding to S protein RBD and mutated  RBD. In the bottom column labels, H indicates the heavy chain of REGN10987. L indicates the light chain of RENG10933. {BFE change range is from -2.47kcal/mol to 2.98kcal/mol.}}
	\label{fig_RN87_all}
\end{center}
\end{figure}
\begin{figure}[h!]
\setlength{\unitlength}{1cm}
\begin{center}
	\includegraphics[width=6.5in]{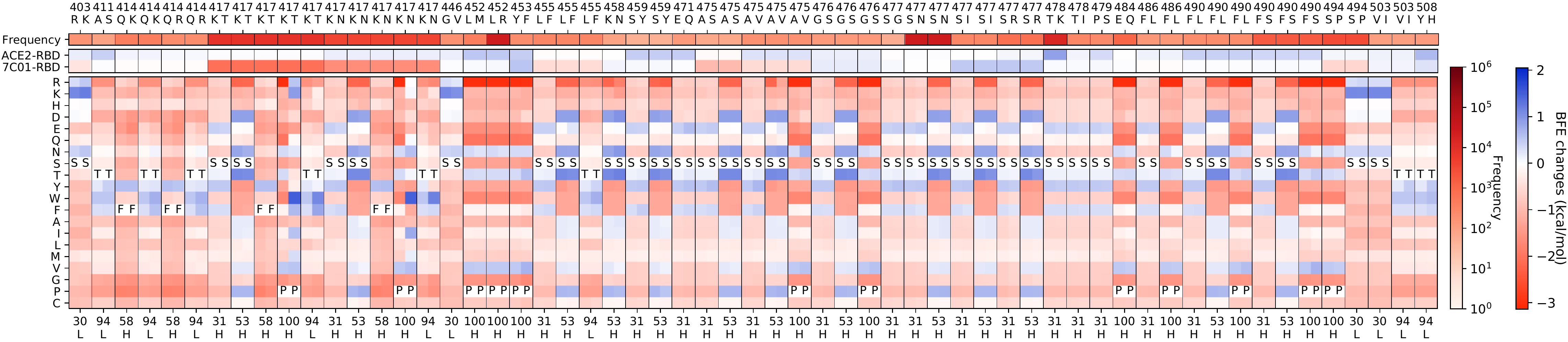}
	\caption{The mutational scanning on antibody LY-CoV016 binding to S protein RBD and mutated  RBD. In the bottom column labels, H indicates the heavy chain of LY-CoV016. L indicates the light chain of LY-CoV016. {BFE change range is from -3.08kcal/mol to 2.03kcal/mol.}}
	\label{fig_7C01_all}
\end{center}
\end{figure}
\begin{figure}[h!]
\setlength{\unitlength}{1cm}
\begin{center}
	\includegraphics[width=1.5in]{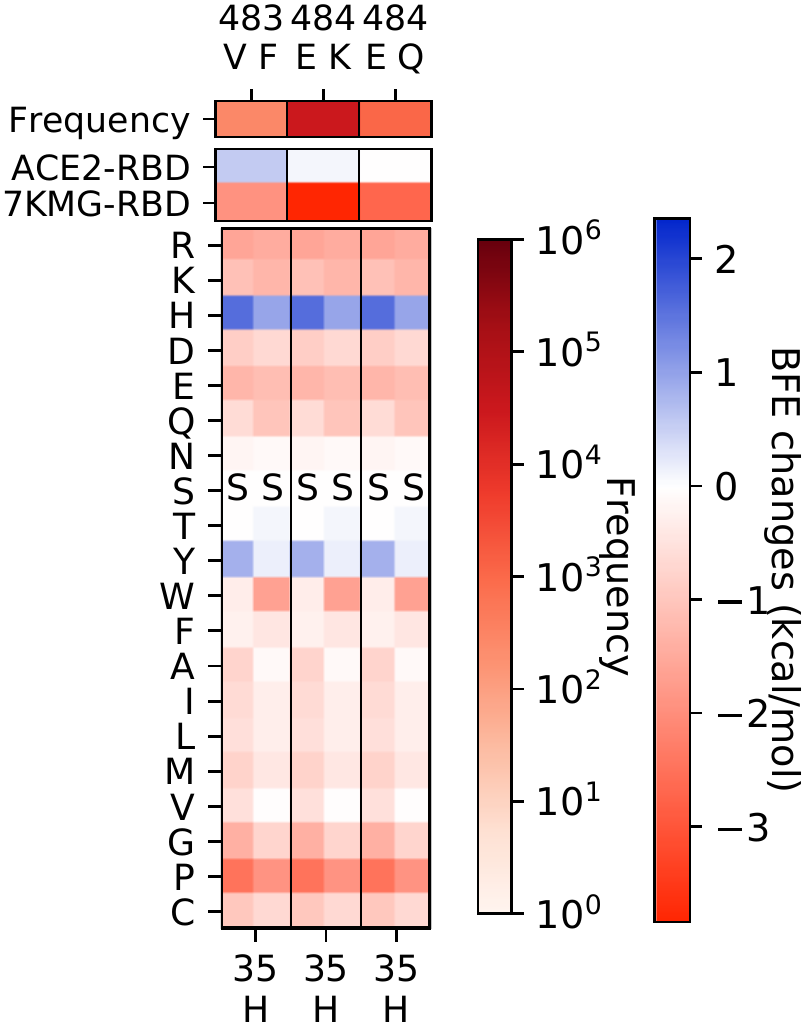}
	\caption{The mutational scanning on antibody LY-CoV555 binding to S protein RBD and mutated  RBD. In the bottom column labels, H indicates the heavy chain of REGN10987. L indicates the light chain of LY-CoV555.  {BFE change range is from -3.79kcal/mol to 2.23kcal/mol.}}
	\label{fig_7KMG_all}
\end{center}
\end{figure}
\begin{figure}[h!]
\setlength{\unitlength}{1cm}
\begin{center}
	\includegraphics[width=6.5in]{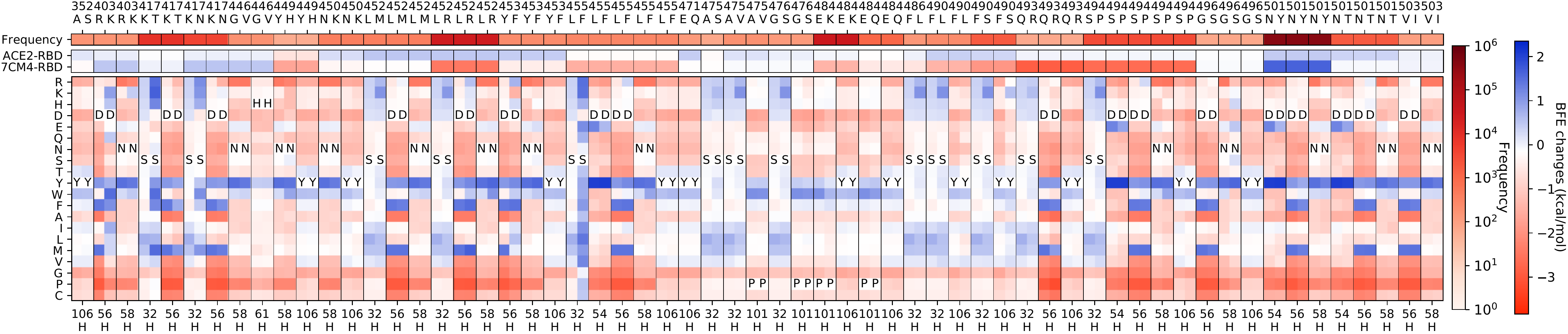}
	\caption{The mutational scanning on antibody CT-P59 binding to S protein RBD and mutated  RBD. In the bottom column labels, H indicates the heavy chain of CT-P59. L indicates the light chain of LY-CoV555.  {BFE change range is from -3.80kcal/mol to 2.26kcal/mol.}}
	\label{fig_7CM4_all}
\end{center}
\end{figure}
Second, this appendix provides the full results of the mutational scanning on antibodies REGN10933 (see Figure~\ref{fig_RN33_all}), REGN10987 (see Figure~\ref{fig_RN87_all}), LY-CoV016 (see Figure~\ref{fig_7C01_all}), LY-CoV555 (see Figure~\ref{fig_7KMG_all}), and CT-P59 (see Figure~\ref{fig_7CM4_all}) binding to S protein RBD and mutated RBD. Note that some of these results do not show a good option in designing mutation-proof antibodies. For example, Figure~\ref{fig_RN33_all} shows that the deep mutational scanning is on antibody REGN10933 of the binding to S protein RBD and mutated RBD. The mutations on S protein RBD are selected from the twelve-selected mutations. As a global observation, an antibody mutation from others to residue tryptophan (denoted as W) favors the binding of S protein RBD and mutated RBD than the others. This provides potential suggestions on antibody design that increasing   tryptophan  populations on the binding interface will enhance the binding affinity of REGN10933 to S protein RBD.

\section*{Competing Interests}
The authors declare no competing interests.

\end{document}